\documentclass[journal=nalefd,manuscript=article]{achemso}
\usepackage[version=3]{mhchem} 
\usepackage{amssymb}
\usepackage{subeqnarray}
\newcommand{\ket}[1]{|#1\rangle}
\newcommand{\bra}[1]{\langle #1|}
\usepackage{color}

\author{Chengwang Niu$^1$}
\email{c.niu@fz-juelich.de}
\author{Patrick M. Buhl$^1$}
\author{Hongbin Zhang$^2$}
\author{Gustav Bihlmayer$^1$}
\author{Daniel Wortmann$^1$} 
\author{Stefan Bl\"{u}gel$^1$}
\author{Yuriy Mokrousov$^1$}
\affiliation
{$^1$Peter Gr\"{u}nberg Institut and Institute for Advanced Simulation, Forschungszentrum J\"{u}lich and JARA, 52425 J\"{u}lich, Germany\\
$^2$Institute of Materials Science, Technische Universit$\ddot{a}$t Darmstadt, 64287 Darmstadt, Germany}	

\title {Topological Nodal-line Semimetals in Two Dimensions with time-reversal symmetry breaking}

\begin{document}


\begin{abstract}
Topological nodal-line semimetals (TNLSs) exhibit exotic physical phenomena due to a one-dimensional (1D) band touching line, rather than discrete (Dirac or Weyl) points. 
While so far proposed two-dimensional (2D) TNLSs possess closed nodal lines (NLs) only when spin-orbit coupling (SOC) is neglected, here using Na$_3$Bi trilayers as an example, we show that 2D TNLSs can been obtained from topological (crystalline) insulators (TI/TCI) by time-reversal symmetry breaking even in the presence of SOC. We further reveal that these obtained NLs are protected by crystalline mirror symmetry, while a mirror symmetry breaking perturbation opens a full gap thus giving rise to a phase transition from 2D TNLS to a quantum anomalous Hall insulator (QAHI). We thereby uncover a close correlation between various topological phases. Remarkably, a strong spin Hall effect, important for transport applications, is predicted in 2D TNLS. Finally, a Na$_2$CrBi trilayer is proposed to realize the 2D TNLS without extrinsic magnetic field. Our work not only proposes a new strategy for realizing 2D TNLSs with truely closed NLs, but also reveals potential applications of TNLS in spintronics.
\end{abstract}

\bigskip

Currently, topological states of quantum matter are discovered in both nontrivial insulators~\cite{Hasan,Qi,Ando} and nontrivial semimetals~\cite{Wan,Wang1,Burkovprb}. In nontrivial insulators, such as topological insulators (TIs) and topological crystalline insulators (TCIs), spin-orbit coupling (SOC) opens up a bulk energy gap with gapless surface/edge states, i.e., Dirac points, in it. Since their discovery, TIs and TCIs have been intensively investigated in three dimensional (3D) and two dimensional (2D) materials. 2D systems bear great potential for the study of exotic phenomena not available in 3D, such as the quantum spin or quantum anomalous Hall effect~\cite{Bernevig,Yu,Fang}. In nontrivial semimetals, conduction and valence bands can cross each other either at zero-dimensional isolated points or along one-dimensional closed curves in so-called Dirac/Weyl semimetals or topological nodal-line semimetals (TNLSs), respectively. Recently, symmetry protected 2D nontrivial semimetals have been predicted theoretically~\cite{Young}, and material realization of the 2D TNLSs are proposed in Hg$_3$As$_2$~\cite{Lu}, PdS~\cite{Jin}, and Ca$_2$As~\cite{Niu} thin films. However, the nodal-line (NL) appears only when SOC is ignored. Similar to graphene~\cite{Yao}, a tiny gap in order of meV opens at the crossing point when SOC is included, thus, strictly speaking, there is no NL in the previously proposed 2D systems ~\cite{Lu,Jin,Niu}. On the other hand in 3D, NLs are obtained even with SOC~\cite{ChenY,Bian,Bian2}. This inspired us to look for the possible realizations  of 2D TNLSs with SOC, and to explore their manifestations, such as~e.g.~the spin Hall effect (SHE), in praticular relevant within the scope
of spintronics apllications~\cite{Sinova}.

\begin{figure}
\centering
\includegraphics{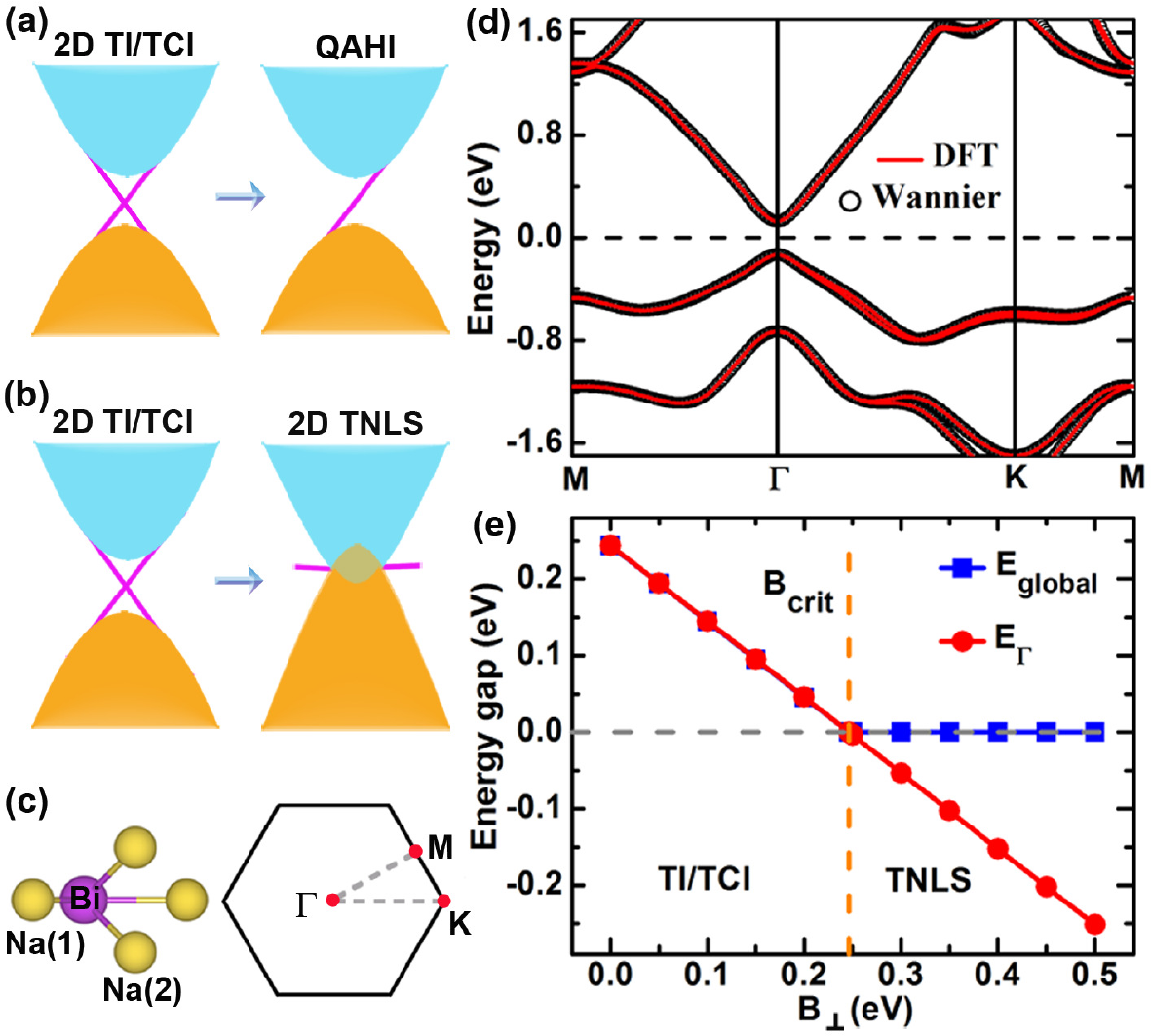}
\caption{ 
Time-reversal symmetry breaking induced topological phase transitions from 2D TI/TCI to (a) QAHI and (b) TNLS. A bulk energy gap is maintained for QAHI while it disappears for TNLS. (c) Side view and Brillouin zone of the 2D Na$_3$Bi trilayer. (d) Comparison between first-principles and Wannier fitted band structures with SOC. (e) Energy gap at the $\Gamma$ point (E$_{\rm \Gamma}$) and the global energy gap (E$_{\rm global}$) of Na$_3$Bi trilayer with
respect to exchange field $B_{\perp}$, showing a phase transition accompanied by a gap closing and reopening at $\Gamma$ point.}
\label{fig1}
\end{figure}

Magnetism may significantly change the electronic and topological properties of the nontrivial materials, such as the creation of Weyl nodes from quadratic bands via the application of magnetic fields as observed in half-Heusler GdPtBi in a recent experiment~\cite{Hirschberger}. For 2D TI and/or TCI, time-reversal symmetry breaking and ferromagnetism can be induced by magnetic doping~\cite{Liucx,Yu,Chang,Wangzf,Zhang,niutci} or chemical decoration~\cite{zhangprl,Wu,Qiao}, and can even result in a phase transition from a 2D TI/TCI to a quantum anomalous Hall insulator (QAHI) (see Figure~\ref{fig1}a). The latter has recently been studied extensively~\cite{Liucx,Yu,Chang,Wangzf,Zhang,niutci,zhangprl,Wu,Qiao} and the QAH effect has been achieved experimentally in Cr-doped (Bi,Sb)$_2$Te$_3$~\cite{Chang}. In QAHI, a bulk energy gap survives. It is thus natural to ask whether there is a magnetically induced metallic transition state in between the 2D TI/TCI and QAHI phases, as the parameters in the system varied. Here we theoretically explore and demonstrate the emergence of a nodal line semimetal as such a transition state with respect to tuning the exchange field, as shown in Figure~\ref{fig1}b.  

Bulk Na$_3$Bi in the hexagonal $P6_3/mmc$ structure (Na$_3$As type) has been theoretically proposed~\cite{Wang1} and experimentally identified~\cite{Liuz} as a topological Dirac semimetal. The bulk crystal structure consists of stacked trilayers along the $c$-axis direction. High quality epitaxial Na$_3$Bi thin films in $c$-axis direction have been fabricated in experiment~\cite{Hellerstedt}, and the triple layers are predicted to be a nontrivial insulator with coexistence of 2D TI and TCI phases~\cite{niuna3bi}. In the present study, taking Na$_3$Bi trilayer as an example, we report the realization of 2D TNLS even with SOC, which is ensured by the crystal mirror symmetry. The spin Hall effect of the 2D TNLS is further investigated.  

Based on density functional calculations as implemented in the film version of the \texttt{FLEUR} code~\cite{fleur,Krakauer}, the maximally localized Wannier functions (MLWFs) are constructed using the {\tt wannier90} code~\cite{Mostofi,Freimuth}. The generalized gradient approximation (GGA) of Perdew-Burke-Ernzerhof (PBE)~\cite{Perdew} is used for the exchange correlation potential during the density functional calculations. SOC is included in the calculations self-consistently. Using a description in terms of MLWFs,  we construct the tight-binding Hamiltonian of the 2D systems and compute the matrix elements of the Pauli matrices $\sigma_{\alpha} (\alpha = x, y, z)$ to consider the effect of the magnetism along different directions, such as out-of-plane ($\sigma_{z}\cdot B_{\perp}$) and in-plane ($\sigma_{x}\cdot B_{\parallel}$). 

Figure~\ref{fig1}c shows a side view of Na$_3$Bi trilayer and the 2D Brillouin zone. There are four atoms in the unit cell, one Bi in Wyckoff 2c position, one Na(1) in 2b position, and two Na(2) in 4f position. Unlike the predicted 2D TNLSs Hg$_3$As$_2$~\cite{Lu}, PdS~\cite{Jin}, and Ca$_2$As~\cite{Niu}, in the Na$_3$Bi trilayer the inversion symmetry is broken while the mirror symmetry $z\rightarrow -z$ and $C_3$ rotation symmetry are preserved. Symmetry analysis shows that the realization of NLs is allowed~\cite{Young}. However, the Na$_3$Bi trilayer is a nontrivial insulator with a dual (TI and TCI) character~\cite{niuna3bi}. The calculated band structure in the presence of SOC is shown in Figure~\ref{fig1}d, from which one can clearly see the energy gap of 0.31 eV at the $\Gamma$ point. 

\begin{figure}[!ht]
\centering
\includegraphics{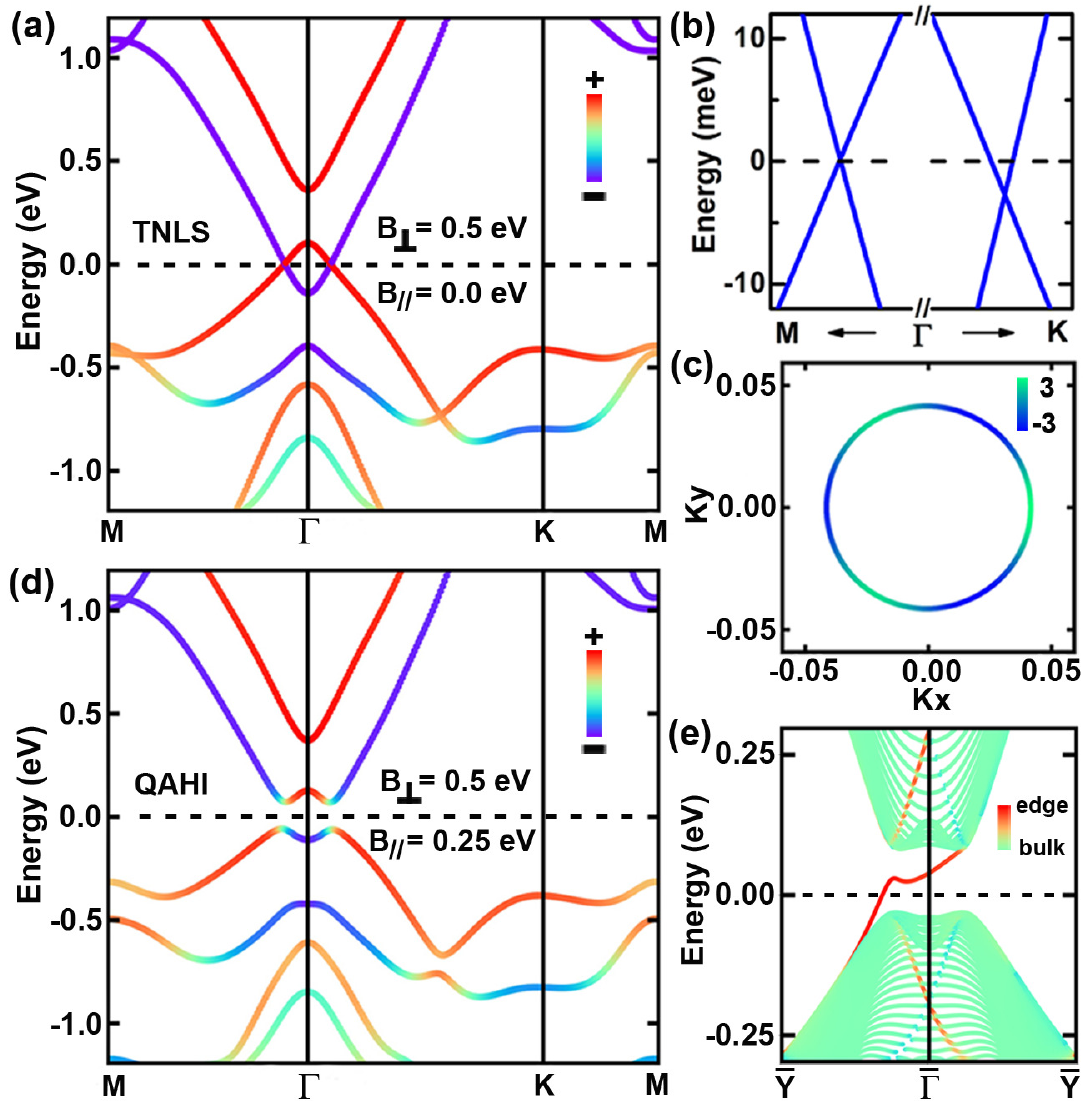}
\caption{ 
Spin-resolved band structures of Na$_3$Bi trilayer with SOC in (a) TNLS phase and (d) QAHI phase under different exchange field B. Colors from purple to red represent the expectation value of $\sigma_z$ indicating the spin polarization. (b) Zoomed-in band structure in the regions of band crossing near the Fermi level. (c) Projected nodal-line in the mirror plane $z = 0$. The color indicated the devation of  corresponding energies from the Fermi energy in meV. (e) Localization-resolved edge states of 1D ribbon with SOC for QAHI phase. The fermi level is set to zero.}
\label{fig2}
\end{figure}

In the following, we explore the magnetically induced effect on Na$_3$Bi trilayer based on the constructed Wannier Hamiltonian. Without a magnetic field, as shown in Figure~\ref{fig1}d, the bands show excellent agreement with the first-principles bands, and the spin-up and spin-down bands are degenerate. Introducing the magnetic field $B_\perp$, which is perpendicular to the mirror plane,  breaks the time-reversal symmetry and splits the spin-degenerate bands. To reveal its influence, we present the variation in the energy gaps, global energy gap E$_{\rm global}$ and energy gap at the $\Gamma$ point E$_{\rm \Gamma}$, with respect to the magnetic field in Figure~\ref{fig1}e. Both E$_{\rm global}$ and E$_{\rm \Gamma}$ are sensitive to the magnetic field as the conduction bands and valence bands approach each other. A closure of the bulk gap at the $\Gamma$ point occurs at $B_\perp$ = 0.246 eV. When further increasing the magnetic field, E$_{\rm global}$ and E$_{\rm \Gamma}$ evolve differently. E$_{\rm \Gamma}$ reopens and increases with an opposite sign while there is no global energy gap, E$_{\rm global}$ remains zero after the critical point, indicating the crossing of conduction and valence bands.

The spin-resolved band structure of Na$_3$Bi trilayer with SOC under $B_\perp$ = 0.5 eV, presented in Figure~\ref{fig2}a, shows the band crossing around the Fermi level clearly. Interestingly, we note that the spin-polarized band crossing encloses the $\Gamma$ point and disperses linearly as shown in Figure~\ref{fig2}b. Taking into account the mirror symmetry $z\rightarrow -z$, the two crossing bands belong to opposite mirror eigenvalues $\pm i$ as they lie in the mirror plane $z = 0$. The band crossing exists not only at isolated points, but also forms a closed NL in the $z = 0$ plane of the Brillouin zone. The projected NL in the mirror plane is plotted in Figure~\ref{fig2}c. Similar to 3D TNLSs~\cite{Kim,Yu2,LiR}, NLs do not appear at a constant energy but have a small energy dispersion. Remarkably, our proposed magnetically induced 2D TNLS in Na$_3$Bi trilayer including SOC is gapless, which is different from currently predicted ones in Hg$_3$As$_2$~\cite{Lu}, PdS~\cite{Jin}, and Ca$_2$As~\cite{Niu}, where SOC opens a tiny energy gap at original crossing points with a negative global gap, so that they are nontrivial semimetals in a strict sense.

\begin{figure}
\centering
\includegraphics{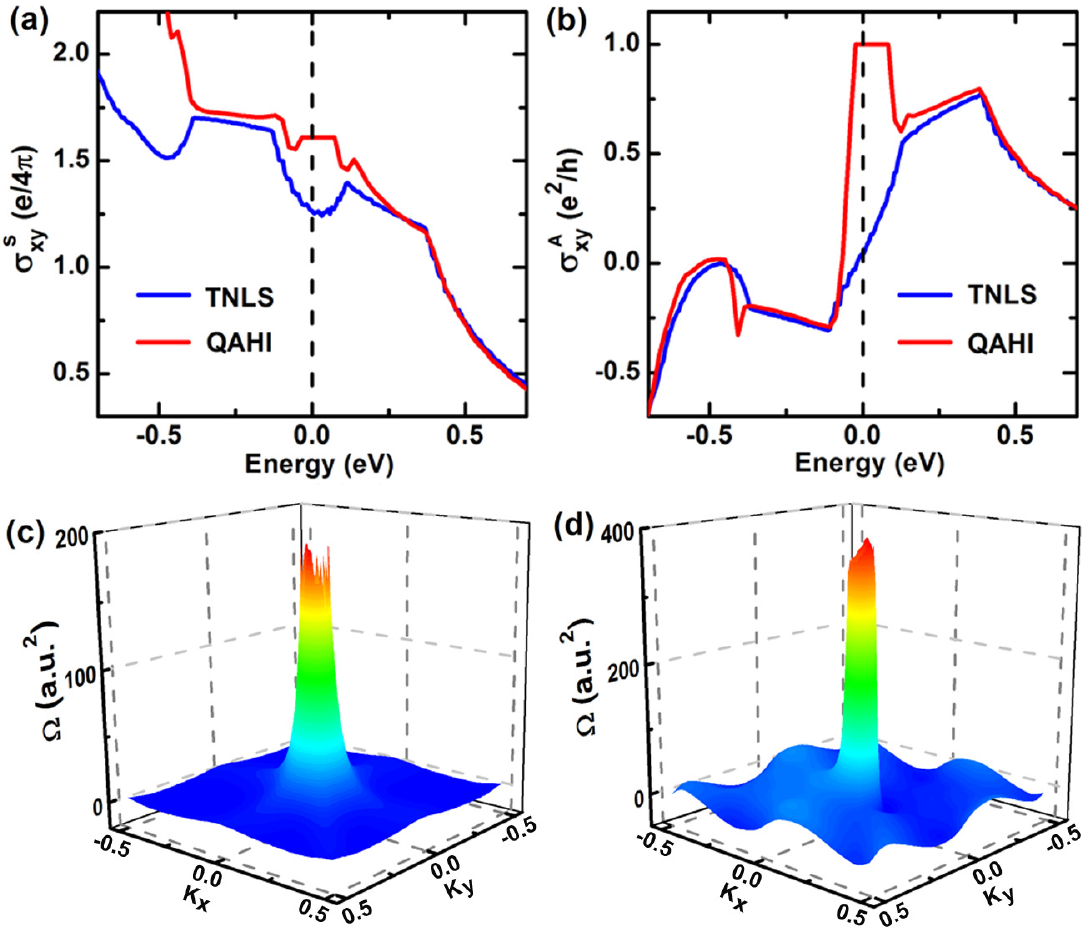}
\caption{ 
(a) Spin Hall conductivity $\sigma_{xy}^{S}$ and (b) anomalous Hall conductivity $\sigma_{xy}^{A}$ as a function of the Fermi level E$_F$ for TNLS phase with $B_\perp$ = 0.5 eV, $B_{\parallel}$ = 0.0 eV and for QAHI phase with $B_\perp$ = 0.5 eV, $B_{\parallel}$ = 0.25 eV. (c) Spin Berry curvature for TNLS phase and (d) Berry curvature for QAHI phase at the Fermi level corresponding to (a) and (b). }
\label{fig3}
\end{figure}

On the other hand, SHE, which is induced by SOC, has generated great interest due to its spintronics applications. The closed NL in the presence of SOC allows us to investigate the SHE in 2D TNLS. Using the fitted Wannier Hamiltonian, we compute the spin Hall conductivity of 2D TNLS, given by the Kubo formula~\cite{Sinova}
\begin{eqnarray}
\sigma_{xy}^{S}=e\hbar\int\frac{d^2k}{(2\pi)^2} \Omega^{S}({\bf k}),
\end{eqnarray}
\begin{eqnarray}
 \Omega^{S}({\bf k})=-2{\rm Im} \sum_{k,m\ne n}\frac{\bra{mk}J_{x}^{s}\ket{nk}\bra{nk}\upsilon_{y}\ket{mk}}{(E_{nk}-E_{mk})^2} .
\end{eqnarray}
Where $ \Omega^{S}({\bf k})$ is the spin Berry curvature summed over occupied valence bands. $J_{x}^{s}=(\hbar/2)\{s_z, \upsilon_x\}$ is the spin current operator with spatial $x$ and spin $s$ component, and $\upsilon_{x/y}$ are the velocity operators. The calculated spin Hall conductivity  $\sigma_{xy}^{S}$ as a function of the Fermi level is presented in Figure~\ref{fig3}a. $\sigma_{xy}^{S}$ is sensitive to the position of the Fermi level and dominated by the large spin Berry curvature $ \Omega^{S}({\bf k})$ contributions around the NL (see Figure~\ref{fig3}c) when the chemical potential is located around crossing points. The emergence of a very large $\sigma_{xy}^{S}$, similar to unexpected $\sigma_{xy}^{S}$ of decorated graphene~\cite{Tuan}, that we observe provides opportunities for practical spintronics applications of 2D TNLS of the type that we study here.

To test the symmetry protection of the 2D NL, firstly we move the two Na(2) atoms by 0.32 \AA~(a large displacement) along in-plane direction, which breaks the $C_3$ rotation symmetry while the mirror symmetry survives. The NL appears under an appropriate magnetic field, indicating that the obtained NL is not a consequence of $C_3$ rotation symmetry. To break the mirror symmetry, the magnetization direction is switched by applying an additional magnetic field parallel to the mirror plane $B_{\parallel}$ on the Na$_3$Bi trilayer. As shown in Figure~\ref{fig2}d, the NL around $\Gamma$ point disappears and an energy gap emerges. In fact, this is similar to the mirror symmetry protected TCIs, where an energy gap in surface/edge states can also be induced by breaking the crystal mirror symmetry~\cite{niutci,Liu}. The induced energy gap is tunable and depends on both $B_\perp$ and $B_{\parallel}$. For $B_\perp$ = 0.5 eV and $B_{\parallel}$ = 0.25 eV, presented in Figure~\ref{fig2}d, the induced energy gap is 0.12 eV, large enough for room temperature application. We then evaluate its anomalous Hall conductivity $\sigma_{xy}^{A}$~\cite{Yao04}, $\sigma_{xy}^{A}=(e^{2}/h)\mathcal C$ (where $\mathcal C$ is the Chern number), and present it as a function of band filling in Figure~\ref{fig3}b.  An integer value of Chern number $\mathcal C = 1$ is indeed acquired when the Fermi level lies inside the energy gap, confirming the QAH effect. This is dramatically different from the nearly zero $\sigma_{xy}^{A}$ in the 2D TNLS phase. Figure~\ref{fig3}d shows the Berry curvature for the whole valence bands in reciprocal space, and the nonzero Berry curvatures are distributed mainly around the original NL near the $\Gamma$ point. The nonzero Chern number is further manifested by the appearance of chiral edge states. Based on the MLWFs, a one-dimensional (1D) nanoribbon with 100-atom width is constructed and considered with applied magnetic field. Apparently, we can see one chiral edge state on each of its edges, Figure~\ref{fig2}e, that is consistent with the calculated Chern number $\mathcal C = 1$. Therefore, the gap opening induced by crystal mirror symmetry breaking signals a phase transition from a 2D TNLS to a QAHI.

\begin{figure}[!ht]
\centering
\includegraphics{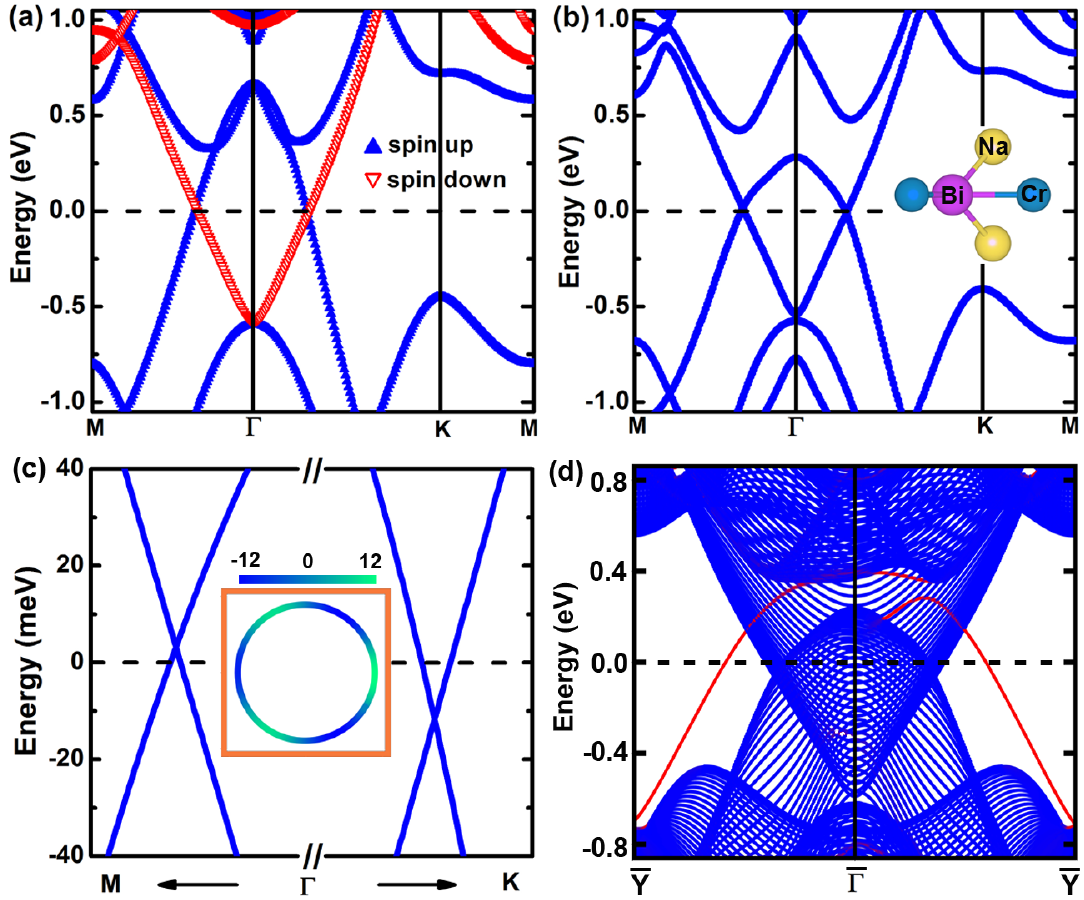}
\caption{ 
Band structures of Na$_2$CrBi trilayers (a) without SOC, where spin up and spin down bands are indicated by solid and empty triangles, respectively, and (b) with SOC. (c) Zoomed-in band structure with SOC in the regions of band crossing near the Fermi level.  (d) Localization-resolved edge states of 1D ribbon with SOC. Color from blue to red represents the weight of atoms located from middle to the edge of 1D ribbon. Inset of (b) and (c ) show side view of the 2D Na$_2$CrBi trilayer and the projected nodal-line in the mirror plane $z = 0$, respectively. }
\label{fig4}
\end{figure}

Importantly, we have further made sure that the NL can be obtained even without extrinsic magnetic field. Taking Na$_3$Bi as an example, we test this possibility by replacing Na(1) by Cr inspired by the ternary compounds with the same structure type as Na$_3$Bi. The mirror symmetry with mirror plane $z = 0$ remains unaltered, as shown in the inset of Figure~\ref{fig4}b. Its ground state is found to be ferromagnetic 
and the magnetic moment is 5\,$\mu_B$ per unit cell. The band structure for the ferromagnetic ground state without SOC is presented in Figure~\ref{fig4}a. The spin-up (solid triangles) and spin-down (empty triangles) bands are spin-polarized and cross each other near the Fermi level. Clearly, a closed NL forms around the $\Gamma$ point. Taking into account the SOC, our magnetic anisotropy calculation shows that the out-of-plane spin state is energetically favorable, which is 5.2 meV lower in energy than the in-plane state. Figure~\ref{fig4}b,c display the band structures with SOC and the zoomed-in views in the regions of band crossing. We observe the clear band crossing around the Fermi level and the bands are found to disperse linearly in the vicinity of the crossings. There is no gap opening for the crossing points even the SOC is tuned on, i.e. the NL remains with SOC. Inset of Figure~\ref{fig4}c shows the projected nodal-line in the mirror plane $z = 0$, that is similar to the nodal line dispersion of the tight-binding analysis but with a larger energy difference. We then focus on the edge states, which is a further confirmation of the nontrivial topological. The tight-binding Hamiltonian of a nanoribbon with a width of 25 unit cells are constructed with the help of Wannier functions of Na$_2$CrBi trilayers. The localization-resolved band structures with Bi-Na termination is presented in Figure~\ref{fig4}d. It is obvious that the edge bands can be clearly distinguished from the bulk state projected ones. There is no SOC-induced gap for the bulk projected bands, agree with the 2D bulk investigations (see Figure~\ref{fig4}b,c), and the nontrivial edge states connect the gapless bulk states.

In summary, we theoretically predicted, for the first time, new family of 2D TNLSs in Na$_3$Bi trilayers. The closed NLs, which rely on the crystalline symmetry $z\rightarrow -z$, survive when SOC is taken into account. As the mirror symmetry is broken for an in-plane magnetic field, the QAHI is obtained with Chern number $\mathcal C = 1$ and exotic edge states. In addition, we predicted very strong SHE in 2D TNLSs, indicating its highly potential applications in spintronics for generating and detecting spin currents.

\bigskip
\noindent\textbf{Notes} 
\\The authors declare no competing financial interest.

\begin{acknowledgement}
This work was supported by the Priority Program 1666 of the German Research Foundation (DFG) and the Virtual Institute for Topological Insulators (VITI). We acknowledge computing time on the supercomputers JUQUEEN and JURECA at J\"{u}lich Supercomputing Centre and JARA-HPC of RWTH Aachen University.
\end{acknowledgement}

\end{document}